\documentclass[12pt,draft]{article}

\usepackage{amssymb,latexsym,pstricks,pst-node}
 
\newtheorem{theorem}{Theorem}[section]

\newcommand{\nc}{\newcommand}
\nc{\stack}[2]{{\begin{array}{c}
\scriptstyle #1 \\ \scriptstyle #2 \end{array}} }
\nc{\C}{{\mathbb C}}
\nc{\R}{{\mathbb R}}
\nc{\HH}{{\mathbb H}}
\nc{\Z}{{\mathbb Z}}
\nc{\dd}{{\rm d}}
\nc{\DD}{{\bf d}}
\nc{\ii}{{\bf i}}
\nc{\jj}{{\bf j}}
\nc{\kk}{{\bf k}}
\nc{\xx}{{\bf x}}
\nc{\pp}{{\bf p}}
\nc{\qq}{{\bf q}}
\nc{\im}{{\rm Im}\:}
\nc{\re}{{\rm Re}\:}
\nc{\ovxi}{\overline{\xi}}
\nc{\ad}{\mathop{\rm ad}\nolimits}
\nc{\tr}{\mathop{\rm tr}\nolimits}
\nc{\su}{{\mathfrak s}{\mathfrak u}(2)}
\nc{\so}{{\mathfrak s}{\mathfrak o} (4)}

\begin{document} 

\title{On Yang--Mills instantons over\\
multi-centered gravitational instantons} 

\author{
G\'abor Etesi
\\ {\it Alfr\'ed R\'enyi Institute of Mathematics,}
\\ {\it Hungarian Academy of Science,} 
\\ {\it  Re\'altanoda u. 13-15, Budapest,}
\\{\it H-1053 Hungary}
\\ {\tt etesi@math-inst.hu}
\\
\\ Tam\'as Hausel
\\ {\it Miller Institute for Basic Research in Science and}\\
{\it Department of Mathematics,}
\\ {\it University of California at Berkeley,}
\\ {\it Berkeley CA 94720, USA}\\
{\tt hausel@math.berkeley.edu}}

\maketitle

\pagestyle{myheadings}
\markright{G. Etesi, T. Hausel: Yang--Mills instantons over
gravitational instantons}

\thispagestyle{empty}

\begin{abstract}
In this paper we explicitly calculate the analogue of the 't Hooft $SU(2)$ 
Yang--Mills instantons on Gibbons--Hawking multi-centered gravitational
instantons, which come in two parallel families: the multi-Eguchi--Hanson,
or $A_k$ ALE gravitational instantons and the multi-Taub--NUT spaces, or
$A_k$ ALF gravitational instantons. We calculate their energy and find the
reducible ones. Following Kronheimer we also exploit the $U(1)$ invariance
of our solutions and study the corresponding explicit 
singular $SU(2)$ magnetic monopole solutions of the Bogomolny
equations on flat $\R^3$.
\end{abstract}

\section{Introduction}
Asymptotically Locally Euclidean or ALE and Asymptotically Locally Flat 
or ALF gravitational instantons are complete, non-compact hyper-K\"ahler
four-manifolds which are intensively studied from both
physical and mathematical sides recently. This paper is a continuation of
the project of constructing $SU(2)$ Yang--Mills instantons on ALF
gravitational instantons started in \cite{ete-hau1} and \cite{ete-hau2}.
In \cite{ete-hau1} we identified all the reducible $SU(2)$ instantons on the 
Euclidean Schwarzschild manifold (which is ALF and Ricci flat, though not
self-dual), and showed that these solutions, albeit not their reducibility, 
were already known to \cite{cha-duf}. Then in \cite{ete-hau2} we went on 
and studied $SU(2)$ instantons on the Taub--NUT space (which is an ALF 
gravitational instanton). Following \cite{jac-noh-reb} we 
exploited the self-duality of the metric, to obtain
a family of $SU(2)$ instantons, which could be considered as the
analogue of the 't Hooft solutions on $\R^4$. (For more historical remarks
the reader is referred to \cite{ete-hau2}.)

Here we carry out the generalization of \cite{ete-hau2} to the more
general case of multi-Taub--NUT spaces (which are also ALF gravitational
instantons). An interesting aspect of our paper is that what we do goes 
almost verbatim in the case of the other family of Gibbons--Hawking
multi-centered gravitational instantons \cite{gib-haw}, namely the 
multi-Eguchi--Hanson spaces.

The only difference in the two cases will be the energy of some of the 
Yang--Mills instantons. Namely in the multi-Eguchi--Hanson case the
infinity may contribute minus a fraction of one to the energy, 
while in the multi-Taub--NUT case the energy is always an integer. 

We will make the calculation of the energy integral in the convenient
framework of considering the $U(1)$-invariant instantons as singular
monopoles on $\R^3$. We will follow \cite{kro} to write down this reduction 
and perform the integral. This way we will also exhibit explicit, but singular,
solutions to the $SU(2)$ Bogomolny equation on $\R^3$. Remarkably we will 
get the same monopoles, regardless of which family of Gibbons--Hawking
metrics we consider: the ALE or ALF.

Then we go on and identify the reducible $U(1)$ instantons among our 
solutions. They are interesting from the point of view of Hodge theory as 
their curvatures are $L^2$ harmonic 2-forms. Namely, only very recently
has the dimension of the space of $L^2$ harmonic forms on ALF gravitational 
instantons been calculated in \cite{hausel-etal}. Remarkably we are able
to identify a basis of generators for these $L^2$ harmonic 2-forms,
arising in this paper as curvatures of reducible 't Hooft instantons. 

In the last part of our paper we put everything together and 
look at the parameter spaces of 
solutions we uncovered in this paper and explain its properties in 
light of the results and point out some future directions to consider.

Our instantons are not new. Most of them were found during the early
eighties by Aragone and Colaiacomo \cite{ara-col} and by Chakrabarti,
Boutaleb-Joutei and Comtet in a series of papers \cite{cha-bou-com}; for a
review cf. \cite{cha}. The 't Hooft solutions and a few more general ADHM
solutions in the Eguchi--Hanson case were written down explicitly in
\cite{bia-fuc-ros-mar}. Finally, $SU(2)$ Yang--Mills instantons over $A_k$
ALE gravitational instantons were classified by Kronheimer and Nakajima
\cite{kro-nak}. 

\section{Instantons over the multi-centered spaces}
In this section we generalize the method of \cite{ete-hau2}, designed for
finding Yang--Mills instantons on Taub--NUT space, to the multi-centered
gravitational instantons $(M_V, g_V)$ of Gibbons and Hawking
\cite{gib-haw}\cite{haw}. 

In our previous paper we used the following result. Suppose $(M,g)$ is a
four-dimensional Riemannian spin-manifold
which is self-dual and has vanishing scalar curvature. Consider
the metric spin connection $\widetilde{\nabla}_S$ of the rescaled manifold
$(M, \tilde{g})$ with $\tilde{g}=f^2g$ where $f$ is harmonic (i.e.,
$\triangle f=0$ with respect to $g$). This $\so$-valued connection lives
on the complex spinor bundle $SM$. Take the $\nabla^-$ component of
$\widetilde{\nabla}_S$. This connection can be constructed as the
projection onto the chiral spinor bundle $S^-M$, according to the
splitting $SM=S^+M\oplus S^-M$ and can be regarded as an $\su^-$-valued
connection. This is because the above splitting of the spinor bundle is
compatible with the Lie algebra decomposition $\so\cong\su^+\oplus\su^-$. 
In light of a result of Atiyah, Hitchin and Singer \cite{ati-hit-sin}
(see also \cite{ete-hau2}) $\nabla^-$ is self-dual with respect to $g$.
These ideas in the case of flat $\R^4$ were first used by Jackiw, Nohl and
Rebbi \cite{jac-noh-reb}, in the case of ALE gravitational instantons in
\cite{cha-bou-com} while for the Taub--NUT case by the authors
\cite{ete-hau2} to construct plenty of new instantons. Our aim is to
repeat this method in the present more general case.

If $\widetilde{\nabla}_S$ is represented locally by an $\so$-valued 1-form 
$\tilde{\omega}$ then we write $A^-$ for the $\su^-$-valued connection
1-form of $\nabla^-$ in this gauge over a chart $U\subset M$. 

Now we turn our attention to a brief description of the
Gibbons--Hawking spaces denoted by $M_V$. This space
topologically can be understood as follows. There is a circle action
on $M_V$ with $k$ fixed points $p_1,\dots,p_k\in M_V$, called
NUTs\footnote{The reason for writing the nut with block letters is the
following. In 1951 Taub discovered an empty space solution of the
Lorentzian Einstein equations \cite{tau} whose maximal analytical
extensions were found by Newman, Unti and Tamburino in 1963 \cite{NUT}.
Hence this solution is referred to as Taub--NUT space-time. 

On the other hand in 1978 Gibbons and Hawking presented a
classification of known gravitational instantons taking into account the 
topology of the critical set of the Killing field appearing in theses
spaces \cite{egu-gil-han} \cite{gib-haw}. Those whose critical set
contains only isolated points were 
called ``nuts'' while another class having two dimensional spheres as
singular sets were named as ``bolts''. It is a funny coincidence that an 
example for the former class is provided by the generalization
of the Riemannian version of Taub--NUT space-time, called multi-Taub--NUT
space in the present work.}. The
quotient is $\R^3$ and we denote the images of the fixed points also by
$p_1,\dots ,p_k\in \R^3$. Then $U_V:=M_V\setminus \{ p_1,\dots, p_k\}$ is
fibered over $Z_V:=\R^3\setminus\{ p_1,\dots,p_k\}$ with $S^1$ fibers. The
degree of this circle bundle around each point $p_i$ is one.  

The metric $g_V$ on $U_V$ looks like (cf. e.g. p. 363 of
\cite{egu-gil-han})
\begin{eqnarray}
\dd s^2=V(\dd x^2+\dd y^2+\dd z^2)+{1\over V}(\dd\tau +\alpha )^2,
\label{metrika}
\end{eqnarray}
where $\tau\in (0,8\pi m]$ parametrizes the circles and
$x=(x,y,z)\in\R^3$; the smooth function $V: Z_V\rightarrow\R$ and the
1-form $\alpha\in C^\infty (\Lambda^1Z_V)$ are defined as follows:
\begin{equation}
V(x ,\tau )=V(x )=c+\sum\limits_{i=1}^k{2m\over\vert x-p_i\vert}
,\:\:\:\:\:\dd\alpha =*_3\dd V.
\label{Vfuggveny}
\end{equation}
Here $c$ is a parameter with values 0 or 1 and $*_3$ refers to the
Hodge-operation with respect to the flat metric on $\R^3$. We can see
that the metric is independent of $\tau$ hence we have a Killing field on 
$(M_V,g_V)$. This Killing field provides the above mentioned
$U(1)$-action. Furthermore it is possible
to show that, despite the apparent singularities in the NUTs, these
metrics extend analytically over the whole $M_V$.

If $c=0$ with $k$ running over the positive integers we find the 
multi-Eguchi--Hanson spaces. If $c=1$ we just recover the multi-Taub--NUT
spaces. In particular if $c=1$ and  $k=1$ then (\ref{metrika}) is the
Taub--NUT geometry on $\R^4$ (cf. Eq. (6) of \cite{ete-hau2}): this is
easily seen by using the coordinate transformation $x^2+y^2+z^2=(r-m)^2$.
Note that under this transform $V$ has the form (by putting the only NUT
into the origin $r=m$) 
\[V(r)=1+{2m\over r-m}\]
i.e., coincides with the scaling function $f$ found in the Taub--NUT case
(cf. Eq. (8) in \cite{ete-hau2}) with $\lambda =2m$.
From here we guess that in the general case the right scaling functions will
have the shape
\begin{eqnarray} f(x)=\lambda_0+\sum\limits_{i=1}^k{\lambda_i\over\vert
x -p_i\vert}\label{ffuggveny}
\end{eqnarray}
where by an inessential rescaling we can always assume that $\lambda_0$ is
either 0 or 1.

We can prove that these are indeed harmonic functions. 
In order to put our formulas in the simplest form, we introduce the
notation $(x,y,z)=(x^1,x^2,x^3)$ and will use Einstein summation
convention. 

First note that from the form of the metric, we have the following
straightforward orthonormal tetrad of $1$-forms on $U_V$:
\begin{equation}
\xi^0={1\over
\sqrt{V}}(\dd\tau+\alpha),\:\:\:\:\:\xi^1=\sqrt{V}
\dd x^1,\:\:\:\:\:\xi^2=\sqrt{V}\dd x^2,\:\:\:\:\:\xi^3=\sqrt{V}\dd x^3.
\label{tetrad}
\end{equation}
The orientation is fixed such that
$*(\xi^0\wedge\xi^1\wedge\xi^2\wedge\xi^3)=1$ which is the {\it
opposite} to the orientation induced by any of the complex structures 
in the hyper-K\"ahler family.  Let
us thus take a general $U(1)$-invariant function
$f:M_V\rightarrow\R$. It means that $f=f(x^1,x^2,x^3)$ does not depend on
$\tau$. Then we have on $U_V$: 
\[ \dd f={\partial f\over \partial x^i}\dd x^i={1\over\sqrt{V}}{\partial
f\over \partial x^i}\:\xi^i\:\:\:\:\:(i=1,2,3).\] 
By using the above orthonormal tetrad
(\ref{tetrad}), we see that 
\[*\dd f=\varepsilon^i_{\:\:jk}{1\over\sqrt{V}} {\partial f\over \partial
x^i}\:\xi^j\wedge\xi^k \wedge\xi^0=\varepsilon^i_{\:\:jk}{\partial f\over
\partial x^i}\:\dd x^j\wedge\dd x^k\wedge (\dd\tau +\alpha ).\] 
(Note that $*$ is the Hodge star operator on $(M_V,g_V)$.)
Consequently 
\[\triangle f= \delta\dd f=-*\dd *\dd f={\partial^2 f\over\partial
x^2}+{\partial^2 f\over \partial y^2}+{\partial^2 f\over \partial z^2}. \] 
Thus we see that the $U(1)$-invariant $f$ is harmonic on $(M_V,g_V)$ if
and only if it is harmonic on the flat $\R^3$. Positive 
harmonic functions on $\R^3$ which are bounded at infinity and have
finitely many point-singularities, with at most inverse polynomial growth, 
have the shape  \[f(x )=\lambda_0+\sum_i {\lambda_i \over |x -q_i|},\] where
$q_i$'s are finitely many points in $\R^3$. 
Note if we want our function $f:M_V\rightarrow\R$ to be a harmonic
function with only point singularities, we need to place the $q_i$'s
at the NUTs $p_i$ of the metric. 
Thus we have found  all reasonable positive $U(1)$-invariant, harmonic
functions  on $(M_V,g_V)$ with point singularities and bounded at
infinity. They are of the form (\ref{ffuggveny}).

Now we determine the Levi--Civit\'a connection of the re-scaled metric
$\tilde{g}=f^2g_V$ restricted to $U_V$. By
using the trivialization (\ref{tetrad}) of the tangent bundle $TU_V$, the
Levi--Civit\'a connection can be represented by an $\so$-valued 1-form
$\tilde{\omega}$ on $U_V$. With the help of the Cartan equation we can
write
\[\dd\tilde{\xi}^i=-\tilde{\omega}^i_j\wedge\tilde{\xi}^j.\]
Taking into account that $\tilde{\xi}^i=f\xi^i$, this yields
\[\dd\xi^i+\dd (\log f)\wedge\xi^i= -\tilde{\omega}^i_j\wedge\xi^j.\]
As we have seen, $f$ does not depend on $\tau$, therefore we have an
expansion like
\[\dd (\log f)={1\over{\sqrt V}}\:{\partial\log f\over \partial
x^j}\:\xi^j\:\:\:\:\:(j=1,2,3).\]
Putting this and $\dd\xi^i=-\omega^i_j\wedge\xi^j$ for the original
connection into the previous Cartan equation, we get
\[\left(\omega^i_j +{1\over \sqrt{V}}\:{\partial\log f\over \partial
x^j}\:\xi^i\right)\wedge\xi^j=\tilde{\omega}^i_j\wedge\xi^j,\]
consequently the local components of the new connection on $U_V$, after
antisymmetrizing, have the shape
\[\tilde{\omega}^i_j=\omega^i_j+{1\over\sqrt{V}}\left({\partial\log
f\over\partial x^j}\:\xi^i-{\partial\log f\over\partial
x^i}\:\xi^j\right) .\]
(Here it is understood that $\partial\log f/\partial x^0 =0$.) 
By the aid of the Cartan equation in the original metric, the
components of the original connection $\omega$
take the form on $U_V$ 
\[\omega^1_2=-{1\over 2\sqrt{V}} {\partial \log V \over \partial x^3
}\xi^0 +
            {1\over 2\sqrt{V}} {\partial\log V\over \partial x^2}\xi^1 -
           {1\over 2\sqrt{V}} {\partial\log V \over \partial x^1}\xi^2\]
and \[\omega^0_1=-{1\over 2\sqrt{V}}{\partial\log V \over \partial
x^1}\xi^0 +
            {1\over 2\sqrt{V}} {\partial\log V \over \partial x^3}\xi^2 -
           {1\over 2\sqrt{V}} {\partial\log V \over \partial x^2}\xi^3,\]
and cyclically for the rest. Notice that from this explicit form we see
that in this gauge the Levi--Civit\'a connection is itself self-dual i.e., 
$\omega^0_1=\omega^2_3$ etc. (cf. p. 363 of 
\cite{egu-gil-han}). Consequently it cancels out if we project
$\tilde{\omega}$ onto the $\su^-$ subalgebra via 
\[A^-_{\lambda_0,\dots,\lambda_k} ={1\over
4}\sum\limits_{a=1}^3\sum\limits_{i,j=0}^3\left(\overline{\eta}^i_{a,j}
\tilde{\omega}^i_j\right)\:\overline{\eta}_a\]
where the 't Hooft matrices $\overline{\eta}^i$ are given by: 
\[\overline{\eta}_1=\pmatrix{ 0 & 1 & 0 & 0\cr -1 & 0 & 0 & 0\cr 0 & 0 &
0 & -1\cr 0 & 0 & 1 & 0\cr},\:\:\overline{\eta}_2=\pmatrix{0 & 0 & -1 &
0\cr 0 & 0 & 0 & -1\cr 1& 0 & 0 & 0\cr 0 & 1 & 0 & 0\cr},\:\:
\overline{\eta}_3=\pmatrix{0 & 0 & 0 & -1\cr 0 & 0 & 1 & 0\cr 0 & -1 & 0
& 0\cr 1 & 0 & 0 & 0\cr} .\]
Using the identification $\su^-\cong\im\HH$ via
$(\overline{\eta}_1,\overline{\eta}_2,\overline{\eta}_3) 
\mapsto (\ii,-\jj,-\kk)$ we get for $\nabla^-_{\lambda_0,\dots,\lambda_k}$ 
in the gauge (or bundle trivialization) given by (\ref{tetrad}) on $U_V$
that  
\[A^-_{\lambda_0,\dots ,\lambda_k}=  
{\ii\over 2\sqrt{V}}\left(
{\partial\log f\over\partial x^1}\xi^0-{\partial\log f\over
\partial x^3}\xi^2+{\partial\log f\over\partial x^2}\xi^3\right)+\]
\[{\jj\over 2\sqrt{V}}\left({\partial\log f\over
\partial x^2}\xi^0+{\partial\log f\over\partial x^3}\xi^1-{\partial\log
f\over\partial x^1}\xi^3\right) +\]
\[{\kk\over 2\sqrt{V}}\left({\partial\log f\over\partial
x^3}\xi^0-{\partial\log f\over\partial x^2}\xi^1+{\partial\log 
f\over\partial x^1}\xi^2\right) .\]
This long but very symmetric expression can be written in a quite simple
form as follows. Consider the following quaternion-valued 1-form $\xi$
and the imaginary quaternion $\DD (\log f)$
(we use the symbol ``{\bf d}'' to distinguish it from the real 1-form $\dd
(\log f)$)
\[\xi :=\xi^0+\xi^1\ii +\xi^2\jj +\xi^3\kk ,\:\:\:\:\:\DD (\log
f):={\partial
\log f\over\partial x^1}\ii +{\partial\log f\over\partial x^2}\jj
+{\partial\log f\over\partial x^3}\kk .\]
It is easily checked that with this notation the connection takes the
simple shape
\begin{equation}
A^-_{\lambda_0,\dots ,\lambda_k}=\im{\DD (\log f)\:\xi\over 2\sqrt{V}}.
\label{insztanton}
\end{equation}
This form emphasizes the analogy with the case of flat $\R^4$ (cf. p.
103 of \cite{fre-uhl}). By construction these connections are self-dual
over $(U_V, g_V)$; but we will prove in the next sections that they are
furthermore gauge equivalent to smooth, self-dual connections over the
whole $(M_V,g_V)$ and have finite energy. 

We remark that there is a familiar face in the crowd (\ref{insztanton}). This
is the solution corresponding to the choice $f=V$. The Yang--Mills
instanton (\ref{insztanton}) is then the same as the projection of the
Levi--Civit\'a connection of $M_V$
onto the other chiral bundle $S^+M_V$, which is easy to see from the form
of the Levi--Civit\'a connection calculated earlier in this section. We
denote this connection by $\nabla_{metric}$. This is the solution which we 
called in \cite{ete-hau2} the Pope--Yuille solution of unit
energy in the Taub--NUT case \cite{pop-yui}\cite{cha-bou-com}. In the
Eguchi--Hanson case this is the solution of energy 3/2 found in
\cite{cha-bou-com} and later again in \cite{kim-yoo}. 

To conclude this section, we write down the field strength
or curvature of (\ref{insztanton}) over $U_V$.  The field strength of 
a connection $\nabla$ with connection 1-form $A$ over a chart is $F=\dd
A+A\wedge A$. Therefore we can see by (\ref{insztanton}) that our field
strength has the form over $U_V$
\[F^-_{\lambda_0,\dots,\lambda_k}=-{\dd V\over 4V^{3/2}}\wedge\im
(\DD (\log f)\:\xi )+{1\over 2\sqrt{V}}\im\left(\DD (\log
f)\:\dd\xi\right) +\]
\[{1\over 2\sqrt{V}}\im\left(\dd\DD (\log f)\wedge\xi\right) +{1\over
4V}\left(\im\left(\DD (\log f)\:\xi\right)\wedge\im\left(\DD (\log
f)\:\xi\right)\right) .\]
The terms in the first line can be adjusted as follows. Using the identity
\[\dd\xi = *_3{\dd V\over\sqrt{V}}-{\dd V\over 2V}\wedge\ovxi\]
we can write them in the form
\[\left( -{\dd V\over
2V^{3/2}}\wedge\xi^0+*_3{\dd V\over 2V}\right)\DD (\log f)
=-{\DD (\log f)\over 4V^2}\re\left(\DD V\xi\wedge\ovxi\right)\]
with
\[\DD V={\partial V\over\partial x^1}\ii +{\partial V\over\partial
x^2}\jj+{\partial V\over\partial x^3}\kk .\]
One immediately sees at this point that these two terms are
self-dual with respect to $g_V$ at least over $U_V$ because
$\xi\wedge\ovxi$ is a basis for self-dual 2-forms. Self-duality of the
remaining two terms is not so transparent; however a tedious but
straightforward calculation assures us about it. So we can conclude that
the connections $\nabla^-_{\lambda_0,\dots,\lambda_k}$ are self-dual with
respect to $g_V$ at least over $U_V$. 

The action, or energy, or $L^2$-norm of the connection (if
exists) is the integral
\begin{equation}
\Vert F^-_{\lambda_0,\dots,\lambda_k}\Vert^2={1\over
8\pi^2}\int\limits_{M_V}\vert
F^-_{\lambda_0,\dots,\lambda_k}\vert^2_{g_V}=-{1\over
8\pi^2}\int\limits_{M_V}
\tr\left(F^-_{\lambda_0,\dots,\lambda_k}\wedge
*F^-_{\lambda_0,\dots,\lambda_k}\right) .
\label{energia}
\end{equation}
Next we turn our attention to the extendibility of (\ref{insztanton}) over
the NUTs and its asymptotical behaviour in order to calculate the above
integral.

\section{A gauge transformation around a NUT}

Our next goal is to demonstrate that the self-dual connections just
constructed are well-defined over the whole $M_V$ up to gauge
transformations. As we have seen, the gauge (\ref{insztanton}) contains
only pointlike singularities hence if we could prove that the energy
of $\nabla^-_{\lambda_0,\dots,\lambda_k}$ is finite in a small ball around
a fixed NUT then, in light of the removable singularity theorem of
Uhlenbeck \cite{uhl} we could conclude that our self-dual connections
extend through the NUTs after suitable gauge transformations. 
However the direct calculation of (\ref{energia}) is complicated because
of its implicit character. Consequently it will be performed in the next
section, here we write down a gauge transformation explicitly, such that
the resulting connection will be easily extendible over the NUTs. 
To this end, we derive a useful decomposition of
(\ref{insztanton}).

To keep our expressions as short as possible, we introduce further
notations: let us write $r_j(x):=|x-p_j|$ and 
\[V_i:=c+{2m\over
r_i},\:\:\:\:\:f_i:=\lambda_0+{\lambda_i\over r_i},\]
and define the 1-form $\alpha_i$ on $\R^3$ by the equation
$*_3\dd\alpha_i=\dd V_i.$ With these notations we introduce a new real 
valued function $a_i$ on $U_V$ as follows:
\begin{equation}
V=:a_iV_i.
\label{aifuggveny}
\end{equation}
One easily calculates
\[a_i=1+{2m\over 2m+cr_i}\sum\limits_{j\not= i}{r_i\over r_j}.\]
In the same fashion by putting the fixed NUT $p_i$ into the origin of
$\R^3$ (i.e., $p_i=0$) we can write
\[\DD (\log f)=-{1\over f}(x^1\ii +x^2\jj +x^3\kk
)\sum\limits_{j=1}^k{\lambda_j\over r^3_j}+{1\over f}\sum\limits_{j\not=
i}{\lambda_j\over r^3_j} (p^1_j\ii +p^2_j\jj +p^3_j\kk ).\]
On the other hand,
\[\DD (\log f_i)=-{1\over f_i}{\lambda_i\over r^3_i}(x^1\ii +x^2\jj +
x^3\kk ),\]
therefore we can write for a real valued function $b_i$ on $U_V$ that
\begin{equation}
\DD (\log f) =:b_i\DD (\log f_i)+ \pp_i
\label{bifuggveny}
\end{equation}
where
\[b_i={1+\sum\limits_{j\not= i}{\lambda_j\over\lambda_i}\left({r_i\over
r_j}\right)^3\over 1+\sum\limits_{j\not= i}{\lambda_j\over
\lambda_i+\lambda_0r_i}{r_i\over r_j}}\]
and we have introduced the function $\pp_i : U_V\rightarrow\im\HH$ given
by
\[\pp_i:={1\over f}\sum\limits_{j\not=
i}{\lambda_j\over r^3_j} (p^1_j\ii +p^2_j\jj +p^3_j\kk ).\]

As a next step, we have to examine these new objects around the NUT
$p_i$. It is not difficult to see, via the explicit expressions for
$a_i$, $b_i$ and $\pp_i$ that 
\[\lim\limits_{r_i(x)\rightarrow 0}a_i(x)=1,\:\:\:\:\:
\lim\limits_{r_i(x)\rightarrow 0}b_i(x)=1,\:\:\:\:\:
\lim\limits_{r_i(x)\rightarrow 0}\vert\pp_i (x)\vert_\HH =0.\]
These observations show that in fact $a_i$, $b_i$ and $\pp_i$ are well
defined on $U_V\cup\{ p_i\}$. Now putting decompositions
(\ref{aifuggveny}), (\ref{bifuggveny}) into (\ref{insztanton}) we arrive
at the expression
\begin{equation}
A^-_{\lambda_0,\dots,\lambda_k}=\im{b_i\DD (\log
f_i)\xi\over 2\sqrt{a_iV_i}}+\im{\pp_i\xi\over 2\sqrt{a_iV_i}}.
\label{szetszedes}
\end{equation}
From here one immediately deduces that the first term (which formally
coincides with the original 't Hooft solution on flat $\R^4$ when $c=0$;
and the solution on Taub--NUT space constructed in \cite{ete-hau2} when
$c=1$) is singular  while the last term vanish in the NUT $p_i$. In order
to analise the structure of the singular term more carefully, we
introduce spherical coordinates around $p_i$ i.e., the origin of $\R^3$: 
\[x^1:=r_i\sin\Theta_i\cos\phi_i,\:\:\:\:\:x^2:=r_i\sin\Theta_i\sin\phi_i,
\:\:\:\:\:x^3:=r_i\cos\Theta_i.\]
Here $\Theta_i\in (0,\pi ]$ and $\phi_i\in (0, 2\pi ]$ are the
angles. In this way we rewrite the singular term as 
\[\im{b_i\DD (\log f_i)\xi\over 2\sqrt{a_iV_i}}=
{b_i\lambda_i\over
2(\lambda_0r_i+\lambda_i)(r_i+2m)}\left( -{1\over 
a_i}(\dd\tau +\alpha )+(r_i+2m)\alpha_i\right)\qq_i+\]
\[{b_i\lambda_i\over 2(\lambda_0r_i+\lambda_i)}((\sin\phi_i\ii
-\cos\phi_i\jj)\dd\Theta_i
-\kk\dd\phi_i)\]
where we have introduced the notation for the ``radial'' imaginary
quaternion 
\[\qq_i:=\sin\Theta_i\cos\phi_i\ii +\sin\Theta_i\sin\phi_i\jj
+\cos\Theta_i\kk\]
and have exploited the identity $\alpha_i=\cos\Theta_i\dd\phi_i$ at
several points. Now we can easily see that the above expression is
singular because the quaternion $\qq_i$ is ill-defined in the NUT $p_i$
i.e., in the origin $r_i=0$ (all the other terms involved are regular in
$p_i$). Consequently we are seeking for a gauge transformation which
rotates the quaternion $\qq_i$ into $\kk$, for example. This gauge
transformation cannot be performed continuously over the whole 
$U_V\cong\R^3\setminus\{ p_1,\dots,p_k\}\times S^1$ therefore we introduce
the two open subsets
\[U^+_V:=\left\{ (x^1,x^2,x^3,\tau )\:\vert\: x^3\geq 0, \mbox{ i.e., }
\Theta_i\geq {\pi\over 2}\right\},\:\:\:\:\:U^-_V:=\left\{
(x^1,x^2,x^3,\tau )\:\vert\: x^3\leq 0, \mbox{ i.e., }
\Theta_i\leq {\pi\over 2}\right\} .\]
Now it is not difficult to check that the gauge transformations $g^\pm_i:
U^\pm_V\rightarrow SU(2)$ given by
\[g^\pm_i(\tau, r_i,\Theta_i, \phi_i):={\rm exp}\left(\pm\kk{\phi_i\over
2}\right){\rm exp}\left(-\jj{\Theta_i\over 2}\right){\rm
exp}\left(-\kk{\phi_i\over 2}\right)\]
(here exp: $\su\rightarrow SU(2)$ is the exponential map) indeed rotate
$\qq_i$ into $\kk$. (We remark that this gauge transformation is exactly
the same which was used to identify the Charap--Duff instantons with
Abelian ones over the Euclidean Schwarzschild manifold in
\cite{ete-hau1}.) Notice that exp$(\kk\phi_i)g^-_i=g^+_i$ showing that the
two gauge transformations are related with an Abelian one along the
equatorial plane $x^3=0$ i.e., $\Theta_i =\pi /2$.

We can calculate that (cf. \cite{ete-hau1})
\[g^\pm_i((\sin\phi_i\ii -\cos\phi_i\jj)\dd\Theta_i
-\kk\dd\phi_i)(g^\pm_i)^{-1}=-2g^\pm_i\dd\left(
g^\pm_i\right)^{-1}\mp\kk\dd\phi_i\]
therefore we get for the gauge
transformed connection $B^\pm_{\lambda_0,\dots,\lambda_k} :=g^\pm_i
A^-_{\lambda_0,\dots,\lambda_k} (g^\pm_i)^{-1}+g^\pm_i\dd\left(
g^\pm_i\right)^{-1}$ by using decomposition (\ref{szetszedes}) that
\[B^\pm_{\lambda_0,\dots,\lambda_k} ={b_i\lambda_i\over
2(\lambda_0r_i+\lambda_i)(r_i+2m)}\left( -{1\over 
a_i}(\dd\tau +\alpha )+(r_i+2m)(\mp 1+\cos\Theta_i)\dd\phi_i\right)\kk
+\]
\begin{equation}
\left(1-{b_i\lambda_i\over 
\lambda_0r_i+\lambda_i}\right)g^\pm_i\dd\left( g^\pm_i\right)^{-1}+
g^\pm_i\im{\pp_i\xi\over 2\sqrt{a_iV_i}}\left( g^\pm_i\right)^{-1}.
\label{ujinsztanton}
\end{equation}
Now we have reached the desired result: we can see that, approaching the
NUT $p_i$ from the north (i.e., along a curve whose points obey $x^3>0$)
the terms written in the first line of $B^+_{\lambda_0,\dots,\lambda_k}$
remain regular while the (non-Abelian) terms of the second line vanish if
$r_i=0$. The situation is exactly the same from the south if we use
$B^-_{\lambda_0,\dots,\lambda_k}$. Moreover
$B^+_{\lambda_0,\dots,\lambda_k}$ and $B^-_{\lambda_0,\dots,\lambda_k}$
are related via an Abelian gauge transformation along the equator $x^3=0$.
Consequently in this new gauge our instantons are regular in the
particular NUT $p_i$.

By performing the same transformations around all NUTs $p_1,\dots,p_k$ we
can see that in fact the instantons (\ref{insztanton}) extend smoothly 
across all the NUTs.

\section{Kronheimer's singular monopoles and the energy}

In this section we identify our $U(1)$-invariant instantons over the
Gibbons--Hawking spaces with monopoles over flat $\R^3$ carrying
singularities in the (images of the) NUTs $p_1,\dots,p_k$. This
identification enables us to calculate the energy of our solutions as
well. In this section we will follow Kronheimer's work \cite{kro}.

Remember $S^-M_V$ is an $SU(2)$ vector bundle over $M_V$ and the
 $U(1)$ action can be lifted from $M_V$ to $S^-M_V$. Our instantons
$\nabla^-_{\lambda_0,\dots,\lambda_k}$ are self-dual,
$U(1)$-invariant $SU(2)$-connections on this bundle. If we
choose an $U(1)$-invariant gauge in $S^-M_V$ (for example
(\ref{insztanton}) or (\ref{ujinsztanton}) is suitable) then
$A^-_{\lambda_0,\dots,\lambda_k}$ becomes an $U(1)$-invariant
$\su^-$-valued 1-form which we can uniquely express as
\[A^-_{\lambda_0,\dots,\lambda_k}=\pi^*A-\pi^*\Psi (\dd\tau +\alpha )\]
where $A$ and $\Psi$ are a 1-form and a 0-form on $Z_V$ and $\pi
:U_V\rightarrow Z_V$ is the projection. Dividing $S^-M_V$ by the
$U(1)$-action we obtain a $SU(2)$ vector bundle $E$ over $Z_V$ together
with a pair $(A,\Psi )$ on it. Omitting $\pi^*$ we calculate
\begin{equation}
F^-_{\lambda_0,\dots,\lambda_k}=\nabla
A^-_{\lambda_0,\dots,\lambda_k}=(F-\Psi\dd\alpha )-\nabla\Psi\wedge 
(\dd\tau +\alpha ).
\label{dekompozicio}
\end{equation}
Here we have used the notation $F=\nabla A$. One finds that the
self-duality 
$F^-_{\lambda_0,\dots,\lambda_k}=*F^-_{\lambda_0,\dots,\lambda_k}$ of the
original connection is equivalent to $*_3(F-\Psi\dd\alpha )=V\nabla\Psi$
or $*_3F=\nabla (V\Psi)$ since $\dd\alpha =*_3\dd V$. Putting $\Phi
:=V\Psi$ we can write
\[F=*_3\nabla \Phi\]
which is the Bogomolny equation for the pair $(A, \Phi)$ on
$Z_V$. This shows that the pair $(A, \Phi )$ can be naturally interpreted
as a magnetic vectorpotential and a Higgs field while $F$ as a magnetic
field on $Z_V$. Notice however that in this case the Higgs field $\Phi$ is
singular at the images of the NUTs hence the reason we have to use the
punctured $\R^3$ denoted by $Z_V$.

In our case, by using (\ref{insztanton}) we can write down the pair
$(A,\Phi )$ explicitly. We easily find that
\begin{eqnarray}
\Phi ={\DD (\log f)\over 2},\:\:\:\:\:A =\im{\DD (\log f)\ii\over
2}\dd x^1 +\im{\DD (\log f)\jj\over 2}\dd x^2+\im{\DD (\log f)\kk\over
2}\dd x^3.
\label{monopole}
\end{eqnarray}
In this framework one can find a simple formula for the energy we are
seeking for. In what follows define 
$Z^R_\varepsilon$ to be the intersection of a
large  ball of radius $R$ in $\R^3$ containing all the NUTs and the
complements of  small balls of radius $\varepsilon_i$ surrounding the NUTs
$p_i$. Putting (\ref{dekompozicio}) into (\ref{energia}) we can write 
\[8\pi^2\Vert F^-_{\lambda_0,\dots,\lambda_k}\Vert^2=\int\limits_{M_V}
\left\vert(F-\Psi\dd\alpha )-\nabla\Psi\wedge
(\dd\tau +\alpha
)\right\vert^2_{g_V}=\]
\[-\int\limits_{M_V}V\left({1\over V^2}\vert
F-\Psi\dd\alpha\vert^2+
\vert\nabla\Psi\vert^2\right)\wedge (\dd\tau +\alpha )= 
16\pi m\lim\limits_\stack{R\rightarrow\infty}{\varepsilon_i\rightarrow 
0}\:\:\int\limits_{Z^R_\varepsilon}V\vert\nabla\Psi\vert^2\]
taking into account the Bogomolny equation in the form $F-\Psi\dd\alpha
=*_3V\nabla\Psi$. By writing 
$2V\vert\nabla\Psi\vert^2=\dd *_3\dd (V\vert\Psi\vert^2)$ and exploiting 
Stokes' theorem we arrive at the useful formula 
\[8\pi^2\Vert F^-_{\lambda_0,\dots,\lambda_k}\Vert^2 =8\pi
m\lim\limits_\stack{R\rightarrow\infty}{\varepsilon_i\rightarrow 0}
       \:\:\int\limits_{\partial Z^R_\varepsilon}*_3\dd\left(
\vert\Phi\vert^2\over V\right) .\]
In the above expressions $\vert\:\cdot\:\vert$ is the Euclidean norm on
$\R^3$ but now interpreted as the Killing norm on $\su^-$ i.e.,
it is equal to {\it twice} the usual norm square of a quaternion under the
identification  $\su^-\cong\im\HH$; e.g. $\vert\Phi\vert^2 =
2\vert\Phi\vert^2_{\HH}$. 

In order to determine the exact value of the action of our solutions, we
simply have to calculate the contributions of each components of the
boundary $\partial Z^R_\varepsilon$, in other
words, the NUTs $p_i$ and the infinity of $\R^3$.

First we can see that, using (\ref{aifuggveny}) and (\ref{bifuggveny}), 
for small $\varepsilon_i$ there is an expansion
\[{\vert\Phi\vert^2\over V} ={1\over 2}\left\vert\left(
-{b_i\lambda_i\over\varepsilon_i(\lambda_0\varepsilon_i+\lambda_i)}\qq_i
+\pp_i\right)\sqrt{{\varepsilon_i\over 
a_i(c\varepsilon_i+2m)}}\right\vert^2_\HH =\left\{\begin{array}{ll}
                  1/(4m\varepsilon_i) +O(1) & \mbox{if $\lambda_i\not=0$,}\\
                                  &                             \\   
                  0 & \mbox{if $\lambda_i=0$.}
       \end{array}
\right.\]
This implies that 
\[\left\vert\dd\left({\vert\Phi\vert^2\over
V}\right)\right\vert =\left\{\begin{array}{ll}
                  \vert -1/(4m\varepsilon^2_i) +O(1/\varepsilon_i)\vert &
                                            \mbox{if $\lambda_i\not=0$,}\\ 
                                           &                            \\   
                  0 & \mbox{if $\lambda_i=0$.}
       \end{array}
\right.\]
However in the above integral there is a contribution
$-4\pi\varepsilon_i^2$ by the spheres (the minus sign comes from the 
orientation) consequently each NUT $p_i$ together with the condition
$\lambda_i\not=0$ contributes a factor $8\pi^2$ to the integral i.e., 1 to
the energy. Hence the total contribution is $n$ where $n$ stands for the
number of non-zero $\lambda_i$'s. Clearly $0\leq n\leq k$.

To see the contribution of infinity, we have to understand the
fall-off properties of the function $\vert\Phi\vert^2/V$. Clearly
this is not modified if we put all the NUTs into the origin of $\R^3$. Thus 
asymptotically our functions take the shape
\[V=c+{2mk\over R},\:\:\:\:\: f=\lambda_0+{1\over
R}\sum\limits_{i=1}^k\lambda_i=:\lambda_0+{\lambda\over R}.\]
Putting these expressions into $\vert\Phi\vert^2/V$ and expanding
it into $1/R$ terms one finds the following for large $R$:
\[{|\Phi|^2\over V}={1\over 2}\left\vert -{\lambda\over
R(\lambda_0R+\lambda )}\qq\sqrt{{R\over cR+2mk}}\right\vert^2_\HH =\left\{
          \begin{array}{ll} 
1/(4mkR) + O(1/R^2) & \mbox{if $\lambda_0=0$ and $c=0$,} \\ 
                            &                             \\
O(1/R^2)  & \mbox{otherwise.} 
          \end{array} 
                       \right. \]
Consequently 
\[\left\vert\dd\left({|\Phi|^2\over V}\right)\right\vert =\left\{
            \begin{array}{ll} 
\vert -1/(4mkR^2) + O(1/R^3)\vert & \mbox{if $\lambda_0=0$ and $c=0$,} \\
                 &                                                      \\ 
\vert O(1/R^3)\vert & \mbox{otherwise.} 
            \end{array} 
                                \right. \] 
Since now again $4\pi R^2$ is the volume of the large sphere
(notice that there is no minus sign because of the orientation)
we get that the contribution of infinity is $-8\pi^2/k$ to the above
integral i.e., $-1/k$ to the energy in the case of
the multi-Eguchi--Hanson space $c=0$ with the special limit $\lambda_0=0$.
Otherwise the contribution of infinity is zero.
Also notice that if $n=0$ then $f=\lambda_0$ is a constant hence the
energy is certainly zero.

We summarize our findings in the following

\begin{theorem} The connection $\nabla^-_{\lambda_0,\dots,\lambda_k}$ as
given in (\ref{insztanton}) is a smooth self-dual $SU(2)$ Yang--Mills
connection and has energy $0$ if all $\lambda_i=0$ $(i>0)$ i.e., $n=0$
(in which case $\nabla^-_{\lambda_0,0,\dots,0}$ is the trivial
connection), otherwise we have
\[\Vert F^-_{\lambda_0,\dots,\lambda_k}\Vert^2=\left\{
          \begin{array}{ll}
         n-(1/k) & \mbox{if $\lambda_0=0$, $c=0$,} \\
                            &                             \\
           n & \mbox{otherwise,}
          \end{array}
                       \right. \]
where $k$ refers to the number of NUTs while $n$ is the
number of non-zero $\lambda_i$'s ($i>0$). $\Diamond$ 
\label{energy}
\end{theorem}

\section{The Abelian solutions}

In this section we show that  
(\ref{insztanton}) is reducible to $U(1)$ if and only if for an $i=0,\dots,k$
we have $\lambda_i>0$ (for simplicity we take $\lambda_i=1$), while
 $\lambda_j=0$ if $j\not=i$ and $j=0,\dots,k$.  
First take the case when $\lambda_0\not=0$ but the other $\lambda$'s vanish, 
then our solution (\ref{insztanton}), which we denote by $\nabla_0$, is
trivial. Now suppose that $\lambda_i\not=0$ for $i=1,\dots,k$ but the others 
vanish. Take the NUT $p_i$ and consider the new gauge
(\ref{ujinsztanton}). In the case at hand the second 
line vanishes and (\ref{ujinsztanton}) reduces to the manifestly Abelian
instanton $\nabla_i$ with $B^\pm_i:=B^\pm_{0,\dots,0,1,0,\dots,0}$ (with 1
is in the $i$th place): 
\[B^\pm_i =\left( -{\dd\tau +\alpha \over
Vr_i}+(\mp 1+\cos\Theta_i)\dd\phi_i\right){\kk\over 2} .\]
But in fact these connections are gauge equivalent because
$B^+_i+{1\over 2}\kk\dd\phi_i=B^-_i-{1\over 2}\kk\dd\phi_i$, and
$\nabla_i$ locally can be written as 
\begin{eqnarray} 
B_i=\left( -{\dd\tau +\alpha \over 
Vr_i}+\alpha_i\right){\kk\over 2},
\label{abeli}
\end{eqnarray} 
where $*_3\dd\alpha_i=\dd V_i$. The curvature 
$F_i=\nabla B_i$ of this Abelian connection are the $L^2$ harmonic
$2$-forms found by \cite{ruback} in the multi-Taub--NUT case. 

Now take any reducible $SU(2)$ instanton of the form
(\ref{insztanton}). Then in a suitable gauge it can be brought to the form
\[\sum\limits_i\mu_i B_i ,\] 
where $\mu_i$ are real numbers. This follows by applying a result from
\cite{hausel-etal} which claims that the 2-forms $F_i$ generate the space
of $L^2$ harmonic 2-forms on our space.
Now note that the corresponding Higgs field of this
instanton is $\left(\sum\limits_i {\mu_i\over r_i}\right){\kk\over 2}$.
Since $\vert\Phi\vert$ is gauge invariant we have the following identity
around a particular NUT $p_i$ via (\ref{bifuggveny}) and (\ref{monopole}):
\[|\Phi|^2 ={1\over
2}\left({\mu_i\over\varepsilon_i}+\sum\limits_{j\not=i}{\mu_j\over
r_j}\right)^2={1\over 2}\left\vert
-{b_i\lambda_i\over\varepsilon_i(\lambda_0\varepsilon_i+\lambda_i)}\qq_i
+\pp_i\right\vert^2_\HH .\]
Provided it is not identically zero the right hand side nowhere vanishes. 
We can deduce that the $\mu_i$'s have to be all non-positive or 
non-negative (otherwise $|\Phi|^2$ would be zero at some point).
Without loss of generality we suppose they are all non-negative. Moreover
the right hand side times $\varepsilon_i^2$ tends to either 0 or 1/2 when
$\varepsilon_i$ tends to 0. Therefore we have that $\mu_i$ is either 0 or
1. Moreover for large $R$ we have $1/(2R^2)\sum\limits_i\mu_i$ for the
left hand side while the right hand side asymptotically looks like
\[\vert\Phi\vert^2=\left\{
          \begin{array}{ll}
         1/(2 R^2) +O(1/R^3) & \mbox{if $\lambda_0=0$}, \\
                            &                             \\
         O(1/R^3) & \mbox{if $\lambda_0\not=0$.}
          \end{array}
                       \right. \]
Thus this last expansion implies that $\lambda_0=0$ and
only one $\mu_i=1$, the rest vanishes. This in turn shows that only one
$\lambda_i$ is not zero which proves the following

\begin{theorem} An instanton in the form (\ref{insztanton}) is 
reducible if and only if for an $i=0,\dots,k$ we have  $\lambda_i\not=0$ 
and $\lambda_j=0$ for $j=0,1,\dots,i-1,i+1,\dots,k$; in this case it can
be put into the form (\ref{abeli}). $\Diamond$
\label{abelik}
\end{theorem} 

\section{Conclusion}
In this paper we have explicitly calculated the analogue of the 't Hooft 
$SU(2)$ instantons for multi-centered metrics of $k$ NUTs. 
We found a $k$ parameter family of such solutions 
(parametrized by $\lambda_0,\dots,\lambda_k$ all non-negative numbers
modulo an overall scaling)  one for each NUTs. 
The structure of the space of solutions 
could be best visualized as an intersection of the positive quadrant and 
the unit ball in $\R^k$. The $k$ flat sides of this body 
correspond to the solutions $\lambda_i=0$ for an $i=1,\dots,k$, 
while the spherical boundary component corresponds to $\lambda_0=0$. The
vertices of this body correspond to the reducible solutions, the origin
corresponding to the trivial solution; the rest to the non-Abelian
solutions (\ref{insztanton}). The energy of the solutions are $k$ at the
interior points of the body, while reduces by $1$ for every $\lambda_i=0$
for $i=1,\dots,k$ and by $1/k$ if $\lambda_0=c=0$. In order to see the
boundary solutions as ideal solutions of energy $k$ we can 
imagine an ideal Dirac delta connection of 
energy 1 at each NUT $p_i$ for which $\lambda_i=0$ in the Taub--NUT
case; the situation is the same for the 
multi-Eguchi--Hanson case except that add a further Dirac delta connection
of energy $1/k$ at infinity if $\lambda_0=0$. 

\begin{figure}[here]
\vspace{2cm}
\begin{minipage}[b]{6.5cm}
\pspicture(-2,-.5)(2,2)
\pswedge{3}{0}{90}
\cnode(0,0){2pt}{a}
\cnode(0,3){2pt}{b}
\cnode(3,0){2pt}{c}
\cnode(1.3,1.3){2pt}{d}
\rput(-.7,1.5){$\lambda_1=0$}
\rput(1.4,-.3){$\lambda_2=0$}
\rput(3.3,1.5){$\lambda_0=0$}
\rput(-.2,3.3){$\nabla_2$}
\rput(3.2,-.2){$\nabla_1$}  
\rput(-.2,-.2){$\nabla_0$}
\rput(1.75,1.5){$\nabla_{metric}$}
\endpspicture
\caption{Case of 2-Taub--NUT}
\end{minipage}
\hspace{2cm}
\begin{minipage}[b]{6.5cm}   
\pspicture(-2,-.5)(2,2)
\pswedge{3}{0}{90}   
\cnode(0,0){2pt}{a}    
\cnode(0,3){2pt}{b}  
\cnode(3,0){2pt}{c}
\cnode(2.1213,2.1213){2pt}{d}
\rput(-.7,1.5){$\lambda_1=0$} 
\rput(1.4,-.3){$\lambda_2=0$}
\rput(3.3,1.5){$\lambda_0=0$}
\rput(-.2,3.3){$\nabla_2$}
\rput(3.2,-.2){$\nabla_1$}  
\rput(-.2,-.2){$\nabla_0$}
\rput(2.5,2.4){$\nabla_{metric}$}
\endpspicture
\caption{ Case of Eguchi--Hanson}
\end{minipage}
\end{figure}

A puzzling feature of this description in the multi-Taub--NUT case  
is at the interior of the spherical boundary component of our solution
space, where  $\lambda_0=0$ but no other $\lambda_i$ is zero. These
solutions are not reducible by Theorem~\ref{abelik} but have energy $k$
by Theorem~\ref{energy}. However they are singular points in our solution 
space. Thus either there are other solutions on the other side of
the spherical boundary, or the moduli space of energy $k$ instantons will 
have singularities at non-reducible points, a phenomenon 
which does not occur when the underlying 4-manifold is compact or ALE. The
relation to the multi-Eguchi--Hanson case, explained
in the next paragraph, however might point to the second possibility.  

Following \cite{kro} we have also studied our $U(1)$ invariant instantons as 
singular monopoles on $\R^3$. Interestingly we got the same monopoles 
(\ref{monopole}) from the
two cases ($c=0,1$). This raises the possibility to take a $U(1)$
invariant instanton on a multi-Eguchi--Hanson space (where all the
solutions were classified in \cite{kro-nak}) consider it as a singular
monopole on $\R^3$ and then pull it back to a $U(1)$ invariant instanton on the
corresponding multi-Taub--NUT space. This method might lead to the 
construction of (all) $U(1)$ invariant instantons on a multi-Taub--NUT
space and if so it would
indeed exhibit singular points in the non-reducible part of the moduli 
space as we explained in the paragraph above. We postpone a more thorough  
discussion of these ideas for a future work as well as the construction
of the spectral data \cite{che-kap} \cite{kro} for the singular monopoles. 

\paragraph{\bf Acknowledgement.} The main steps in this work were done
when the first author visited the University of California at Berkeley     
during May 2002. We would like to acknowledge the financial support by
Prof. P. Major (R\'enyi Institute, Hungary) from his OTKA grant No. T26176
and of the Miller Institute for Basic Research in Science at UC Berkeley.

\end{document}